\documentclass{aa}  

\usepackage{longtable}
\usepackage{graphicx}
\usepackage{color}
\usepackage{txfonts}
\begin{document}

   \title{Identifying changing jets through their radio variability}


   \author{I. Liodakis \inst{\ref{inst1}}\thanks{yannis.liodakis@utu.fi}
   \and T. Hovatta \inst{\ref{inst1},\ref{inst4}} \and M. F. Aller \inst{\ref{inst2}} \and H. D. Aller \inst{\ref{inst2}} \and M. A. Gurwell \inst{\ref{inst3}} \and A. L\"ahteenm\"aki \inst{\ref{inst4}, \ref{inst5}} \and M. Tornikoski \inst{\ref{inst4}}}
   \institute{Finnish Center for Astronomy with ESO, University of Turku, Quantum, Vesilinnantie 5, FI-20014, Finland\label{inst1} \and  Aalto University Mets\"ahovi Radio Observatory, Mets\"ahovintie 114,
02540 Kylm\"al\"a, Finland \label{inst4} \and Department of Astronomy, University of Michigan, Ann Arbor, MI 48109-1107 USA \label{inst2} \and Center for Astrophysics | Harvard \& Smithsonian, 60 Garden Street, Cambridge, MA 02138 USA \label{inst3} \and Aalto University Department of Electronics and Nanoengineering,
P.O. BOX 15500, FI-00076 AALTO, Finland \label{inst5}\\}

 
  \abstract
   {Supermassive black holes can launch highly relativistic jets with velocities reaching Lorentz factors as high as $\Gamma>50$. How the jets accelerate to such high velocities and where along the jet do they reach terminal velocity are open questions that are tightly linked to their structure, as well as launching and dissipation mechanisms.}
   {Changes in the beaming factor along the jets could potentially reveal jet acceleration, deceleration, or bending. We aim to (1) quantify the relativistic effects in multiple radio frequencies and (2) study possible jet velocity -- viewing angle variations at parsec scales.}
   {We used the state-of-the-art code {\it Magnetron} to model  light curves from the University of Michigan Radio Observatory and the Mets\"{a}hovi Radio Observatory's monitoring programs in five frequencies covering about 25 years of observations in the 4.8 to 37~GHz range for 61 sources. We supplement our data set with high-frequency radio observations in the 100-340~GHz range from ALMA, CARMA, and SMA. For each frequency we estimate the Doppler factor which we use to quantify possible changes in the relativistic effects along the jets.}
   {The majority of our sources do not show any statistically significant difference in their Doppler factor across frequencies. This is consistent with constant velocity in a conical jet structure, as expected at parsec scales. However, our analysis reveals 17 sources where relativistic beaming changes as a function of frequency. In the majority of cases the Doppler factor increases towards lower frequencies. Only 1253-053 shows the opposite behavior. By exploring their jet properties we find that the jet of 0420-014 is likely bent across the 4.8-340~GHz range. For 0212+735 the jet is likely parabolic, and still accelerating in the 4.8-37~GHz range. We discuss possible interpretations for the trends found in the remaining sources.}
   {}

   \keywords{Physical data and processes: Relativistic processes --  Galaxies: active -- Galaxies: jets }

\maketitle
%

\section{Introduction}

Blazars are a subclass of active galactic nuclei (AGN) with powerful and energetic jets pointed towards our line of sight \citep{Blandford2019}. Due to the alignment of their jets, blazars are among the brightest and most variable sources from low-frequency radio all the way to very high-energy $\gamma$-rays. Their jets show complex structures from the smallest (e.g., \citealp{Kim2020}) to the largest scales (e.g., \citealp{Kharb2010}) with radio emission being an ever present piece of the relativistic jet puzzle. Understanding radio variability on diverse scales can therefore be a potential probe of many aspects of jet microphysics. It is therefore not surprising that several attempts have been made to understand the variability properties of blazars in the radio regime from different perspectives e.g., spectral and multiwavelength  variability, variability amplitudes, and beaming effects etc. (e.g., \citealp{Angelakis2010,Richards2014,Liodakis2017-IV,Liodakis2018-II}). 

Jets in blazars are highly relativistic with Lorentz factors ($\Gamma$) from a few to a few tens (e.g., \citealp{Hovatta2009,Liodakis2015,Liodakis2017-III,Jorstad2017}). Whether the jets are launched with such high $\Gamma$ or are accelerated to the velocities measured at parsec scales is still an open question. Most likely the jets are highly magnetized when launched and are accelerated through the conversion of magnetic to kinetic energy \citep{Vlahakis2004,Vlahakis2004-II,Komissarov2007,Zhang2021}. The jet is confined by either magnetic hoop stress (e.g., \citealp{Spruit1997}) or the pressure of the external medium \citep{Lyubarsky2009,Lyubarsky2010,Liodakis2018-III} into a parabolic shape which favors acceleration. Acceleration continues until the jet magnetization parameter ($\sigma_m$) i.e., the ratio of the magnetic to the kinetic energy flux, becomes $\sigma_m\leq1$ when it likely stops \citep{Vlahakis2003,Vlahakis2004,Lyubarsky2009,Nokhrina2020}. Contrary to observations, which are typically taken in flaring states, relativistic magnetohydrodynamic (RMHD) simulations typically yield $\Gamma$ of only a few (e.g., \citealp{Chatterjee2019}). Very Long Baseline Interferometry (VLBI) observations show that the jet in M87 has a parabolic shape up to the HST-1 region (e.g., \citealp{Biretta1999}) and then transitions to conical \citep{Asada2012}.  Recent observations of other nearby radio galaxies have found similar geometries suggesting this is a common feature of AGN jets \citep{Kovalev2020,Boccardi2021}, while theoretical works predict differences in the location of the transition region between blazar subclasses \citep{Potter2015}. Based on our current understanding of blazar jets (e.g., \citealp{Blandford1979,Marscher1995,Marscher2008}) GHz radio emission should arise at parsec scales where the jets are likely conical and nonaccelerating. However, accelerating jet components found by the MOJAVE program at 15~GHz challenge this interpretation \citep{Homan2009,Homan2015}. Moreover, the ``Doppler crisis'', i.e., the discrepancy between Doppler factors measured in the radio bands and that required to explain the high-energy emission in high-synchrotron peaked sources, has been interpreted  either as the result of decelerating  (e.g., \citealp{Georganopoulos2003-II}) or structured jets (e.g., \citealp{Piner2018}). The Doppler factor ($\delta$) is a function of the velocity and the viewing angle of the jet defined as $\delta=1/(\Gamma[1-\beta\cos\theta])$ where $\theta$ is the viewing angle, $\beta=u/c$ where c is the speed of light, and $\Gamma=1/\sqrt(1-\beta^2)$.

To shed more light on the beaming profiles of the jets, we model the flux-density variations in multiple radio frequencies. Because of synchrotron self-absorption, the emission at different frequencies probes different locations along the jets. Hence, quantifying the relativistic effects as a function of frequency can help us identify changes in the jet velocity or orientation at parsec scales. As flux-density variations occur predominately in the radio cores \citep{Savolainen2002}, this allows us to study sources where their jet structure is unresolved through VLBI. In section \ref{sampl} we discuss the sample and the tools used for the analysis of the radio light curves. In section \ref{d_vs_nu} we explore the relativistic effects as a function of frequency, and in section \ref{og_doppler} we explore the possible origin of the relativistic beaming variations and discuss our results. Our conclusions are presented in section \ref{concl}. Through the paper we have adopted a $\rm \Lambda{CDM}$ cosmological model with $\Omega_m=0.27$, $\Omega_\Lambda=1-\Omega_m$ and $\rm H_0=71$ ${\rm km \, s^{-1} \, Mpc^{-1}}$ \citep{Komatsu2009}.

\section{Data and light-curve modeling}\label{sampl}
\begin{figure*}
\centering
\resizebox{\hsize}{!} {\includegraphics[width=\hsize]{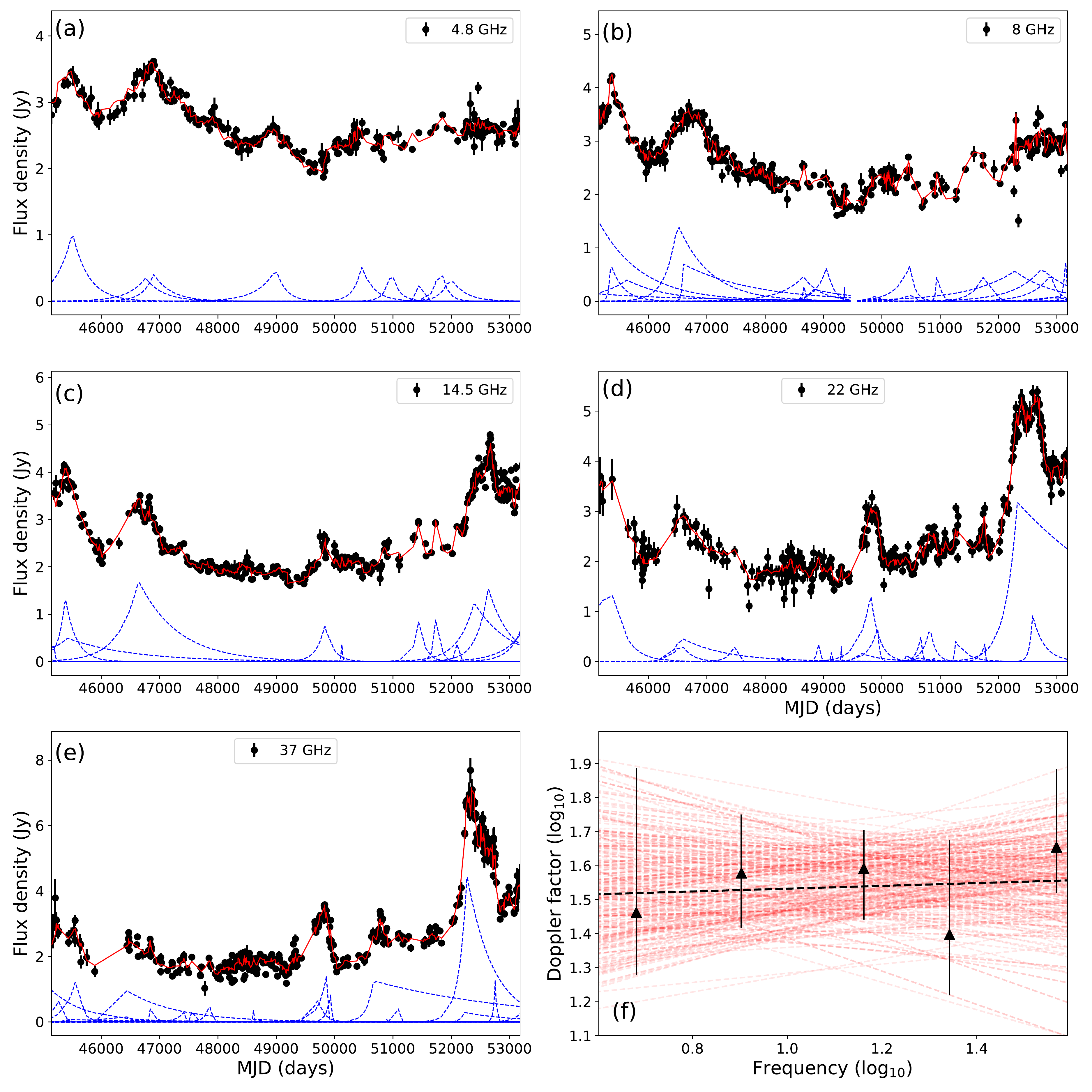}}
\caption{Example of light-curve fitting for 1633+382 for all frequencies (Panel a-e). The red solid line shows the overall fit in each panel, whereas the blue lines show one randomly selected realization of the flares having after subtraction of the background. Here we only consider flares from June 1982 to June 2004. Panel (f) shows the Doppler factor versus frequency relation in log-log space. The black dashed line shows the best-fit line, and the red dashed lines show the uncertainty of the fit by randomly drawing from the joint posterior distribution of the slope (S) and intercept (I).}
\label{plt:doppler_example}
\end{figure*}

We use data from the Mets\"{a}hovi and University of Michigan (UMRAO) Radio observatories for five frequencies: 4.8, 8, 14.5, 22, and 37~GHz covering a few decades of observations \citep{Aller1985,Aller1999,Aller2014,Salonen1987,Terasranta1992,Teraesranta1998,Teraesranta2004,Teraesranta2005}\footnote{https://dept.astro.lsa.umich.edu/datasets/umrao.php}.  Our sample consists of 61 common sources, 35 of which are Flat Spectrum Radio Quasars (FSRQs), 22 are BL Lac objects (BL Lacs), 3 are radio galaxies and one is unidentified. Our earliest observations start in 1965 at 8~GHz and the latest end in 2018 at 14.5 and 37~GHz. The light curves were analyzed using {\it Magnetron}\footnote{\url{https://github.com/dhuppenkothen/magnetronhierarchy}}. Here we provide a brief description, more details can be found in \cite{Huppenkothen2015,Liodakis2018-II}. 

{\it Magnetron} decomposes the light curves into a series of flares superimposed on a Ornstein--Uhlenbeck type stochastic background. Each flare has decoupled exponential rise and exponential decay profiles fully described by four free parameters: position, rise time, amplitude, and flare skewness (decay time/rise time ratio). The parameter space is efficiently explored through diffusive nested sampling \citep{Brewer2009,Brewer2016}. {\it Magnetron} does not deliver a best-fit solution for each source, but rather a posterior distribution of $\sim10^2$ models of flares and backgrounds.  These models represent the different realizations of the flaring activity and underlying stochastic variability in a source given the overall uncertainty in the flare parameters and flare blending, which becomes even more severe at the lower centimeter-band wavelengths. The number of flares is also a free parameter that can vary for different frequencies as shown in Fig. \ref{plt:doppler_example}. Panels (a-d) show one randomly selected realization of the flares in each frequency for 1633+382 (also know as 4C~38.41). Purely stochastic models using a single or a combination of multiple OU-processes have been used on $\gamma$-ray light curves (e.g., \citealp{Sobolewska2014,Burd2021}). However, there is strong evidence that multiwavelength flares, and especially radio flares, are connected to the ejection of jet components (e.g., \citealp{Savolainen2002,Marscher2008,Liodakis2020-II}) suggesting that a combination of flares and stochastic variability, such as the one used by {\it Magnetron}, is a more appropriate model.

For every frequency and for each source, we use the posterior distribution of flares to estimate the distribution of the highest brightness temperature from which we estimate the median and confidence intervals. During the fitting, we use all the available observations for a given source. However, the time period during which a source was observed by the Mets\"{a}hovi and University of Michigan Radio observatories varies not only for each source, but also for each frequency. Hence, it is possible for the light curves of one observatory to include flares not observed by the other. This can introduce a bias when looking for variations of the Doppler factor across frequencies. Therefore, in our analysis we only consider flares during the common observing periods between the observatories. The earliest flare we consider is at MJD~44401 (June 1980) while the latest is at MJD~53179 (June 2004). The variability brightness temperature is estimated as,
\begin{equation}
\rm T_\mathrm{var}=1.47 \cdot 10^{13}\frac{d^2_L \Delta S_\mathrm{ob}(\nu)}{\nu^2t^2_\mathrm{var}(1+z)^4}~K,
\label{eq:tvar_num}
\end{equation}
where $\rm d_L$ is the luminosity distance (Mpc), $\rm \Delta S_\mathrm{ob}(\nu)$ the flare amplitude (Jy), $\nu$ the observing frequency (GHz), $\rm t_\mathrm{var}$ the flare rise time (days), and $\rm z$ is the redshift. We find a wide range of observed brightness temperatures across frequencies. The values range from $\rm 10^8-10^9 K$ all the way to $\rm \sim10^{16} K$, with the median for each frequency to be about $\rm 10^{13}-10^{14} K$. We estimate the Doppler factor by marginalizing over the observed 
$\rm T_\mathrm{var}$ distribution and the Gaussian model for the maximum intrinsic brightness temperature ($\rm T_\mathrm{int,max}$) found in \cite{Liodakis2018-II} with mean $\rm \mu=2.78\times10^{11} K$  and $\sigma=0.72\times10^{11}$ using,
\begin{equation}
\rm \delta_\mathrm{var}=(1+z)\sqrt[3]{\frac{T_\mathrm{var}}{T_\mathrm{int,max}}}.
\label{eq:varia-Doppler}
\end{equation}
This process gives us a distribution of Doppler factors for each source from which we quote the median and 68\% confidence intervals in Table \ref{table:1}. An example of the posterior $\rm \delta_\mathrm{var}$ distribution in all frequencies for 0716+714  is shown in Fig. \ref{plt:doppler_dist} (top panel). The range of Doppler factors is also wide, starting below unity for some radio galaxies, up to almost sixty. The median value across frequencies is $\sim10$ (Fig. \ref{plt:doppler_dist} bottom panel).

\section{Doppler factor versus frequency}\label{d_vs_nu}
  
\begin{figure}
\centering
\resizebox{\hsize}{!} {\includegraphics[width=\hsize]{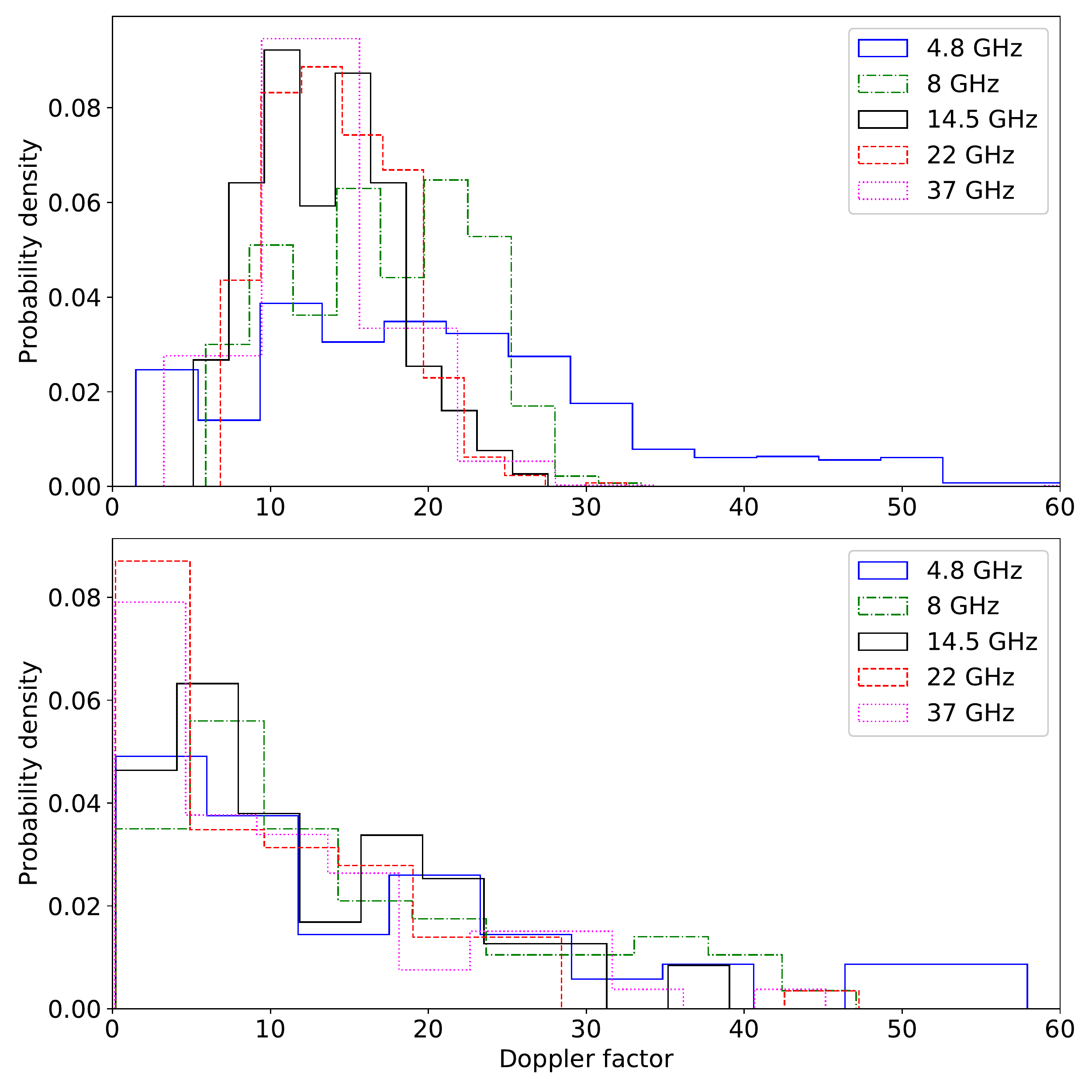}}
\caption{{\bf Top panel:} Posterior Doppler factor  distribution for 0716+714. {\bf Bottom panel:} Median Doppler factor distribution for all frequencies for the sources in our sample. For both panels solid blue is for 4.8, dashed-dotted green for 8, solid black for 14.5, dashed red for 22, and dotted magenta for 37~GHz.}
\label{plt:doppler_dist}
\end{figure}

Figure \ref{plt:doppler_example} (e) shows an example of the Doppler factor versus frequency in log-log space for 1633+382. The resulting posterior distributions for the Doppler factor in individual sources are often asymmetric as shown in Fig. \ref{plt:doppler_dist} (0716+714, top panel). We take that asymmetry into account when trying to statistically establish a persistent Doppler factor versus frequency trend for each source by randomly sampling the posterior $\rm \delta_\mathrm{var}$ distribution for each frequency. We create a new Doppler factor versus frequency relation and then use the Pearson correlation test\footnote{The Pearson correlation test yields a correlation coefficient $
\rho$ defined between [-1,1] where -1 denotes a perfect anti-correlation, 0 no correlation and 1 a perfect correlation. The accompanying p-value is the random chance probability of such correlation. For any p-value>0.05 the correlation is not considered statistically significant.} to estimate the correlation coefficient $\rho$ and the probability (p-value) of a random correlation. We additionally fit a linear model ($\rm y=S*x+I$) in log-log space. We repeat this process $10^3$ times to create the posterior distribution for $\rho$, p-value, $\rm S$, and $\rm I$ from which we estimate median and 68\% confidence intervals. The results of the correlation coefficient, p-values, and best-fit line coefficients are given in Table \ref{table:2}. 
  
Following this procedure, we find that out of the 61 sources only 17 (27.8\%, 10 FSRQs and 7 BL Lacs) show a statistically significant trend of $\rm \delta_\mathrm{var}$ changing with frequency. We discuss our interpretation for the trends below. Based on the Pearson correlation p-values ($P$), we can estimate the false-positive rate, i.e., the number of sources where a significant trend could have been falsely identified,  as $(\sum\limits_{i=1}^{N}{P_i})/N$. We find our false-positive rate to be 18\% (3/17 sources).

\section{Origin of the Doppler factor trend}\label{og_doppler}

Through the analysis discussed above we can identify three different trends:
\begin{itemize}
\item {\bf No statistically significant trend.} This is true for the majority of the sources in our sample  (hereafter Sample A, 44 sources) suggesting no variation of the Doppler factor across frequencies.

\item {\bf Doppler factor increases towards lower frequencies.} This trend is found for 16 out of the 17 sources (hereafter Sample B) that show a statistically significant trend.

\item {\bf Doppler factor increases towards higher frequencies.} This trend is found for only one of the Sample B sources.
\end{itemize}

The fact that the majority of the sources in our sample do not show a statistically significant trend (Sample A) is consistent with the frequently used assumption of a straight conical jet with constant velocity. For the Sample B sources, the most common trend of an increasing  $\rm \delta_\mathrm{var}$ towards lower frequencies has been noted by \cite{Liodakis2017} based on multiwavelength radio observations from the F-GAMMA program \citep{Fuhrmann2016}. Only 1253-055, also known as 3C~279, from Sample B shows the opposite trend i.e., increasing $\rm \delta_\mathrm{var}$ towards higher frequencies. This trend is also confirmed by ALMA observations at 100~GHz (Fig.\ref{plt:highfreq} panel C). The origin of the trends found in Sample B can be attributed to either acceleration or jet bending. We discuss the possible interpretations below. For our comparisons we use the Kolmogorov-Smirnov (K-S) test\footnote{The Kolmogorov-Smirnov test under the null hypothesis that two samples are drawn from the same parent distribution yielding the corresponding probability. We accept that for p-values $>0.05$ we cannot reject the null hypothesis}. We also use the k-sample Anderson-Darling\footnote{The k-sample Anderson-Darling test operates under the same null hypothesis as the K-S test. Similarly, we  do not reject the null hypothesis for p-values $>0.05$. The A-D test is more sensitive to the tails of the distributions, whereas the K-S test is more sensitive to the median values.} (A-D) test to cross-check our results.

\subsection{Changes in the viewing angle}
\begin{figure}
\centering
\resizebox{\hsize}{!} {\includegraphics[width=\hsize]{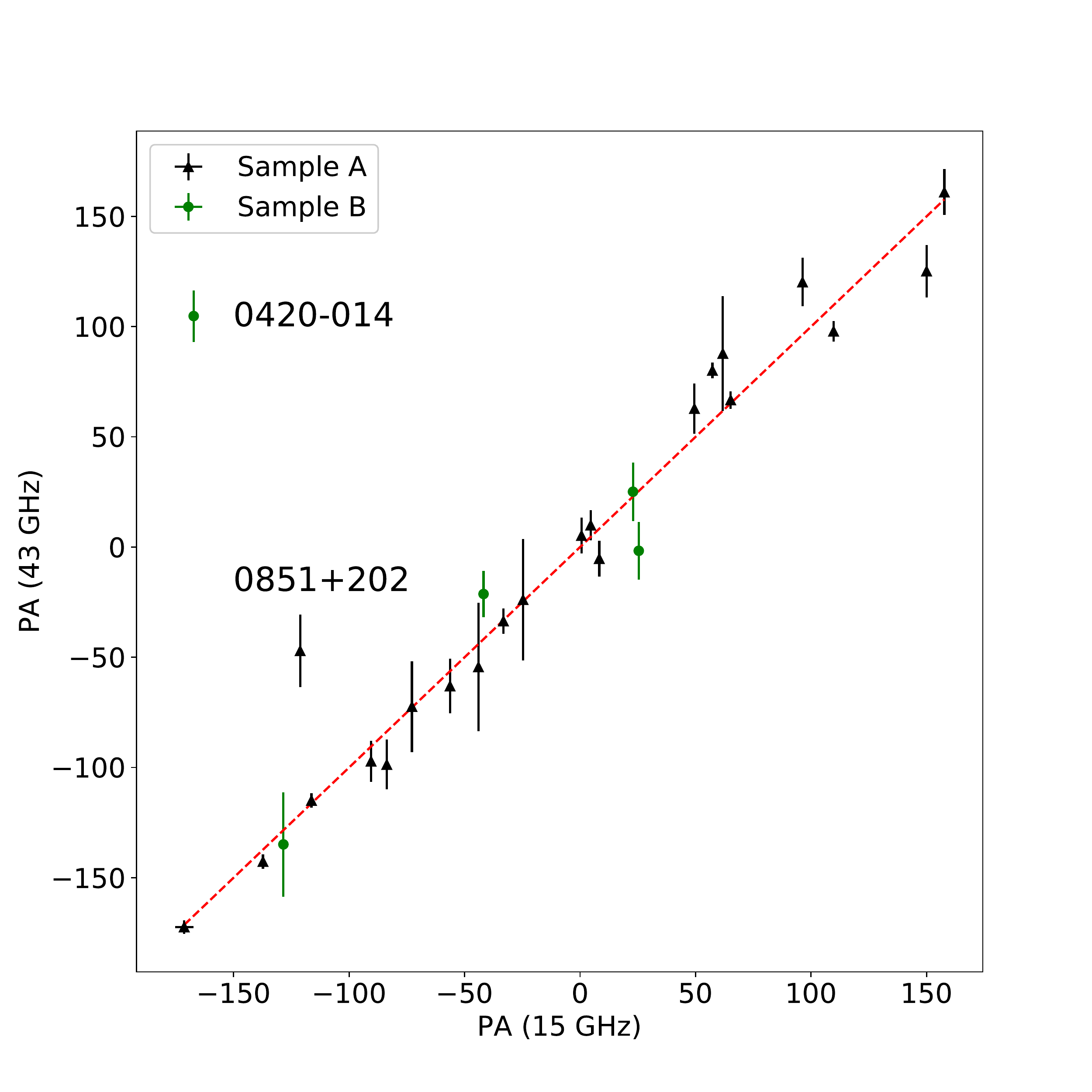}}
\caption{Comparison of the innermost jet position angles at 15 and 43 GHz for Sample A (black) and Sample B (green). The red dashed line shows the 1-1 relation.}
\label{plt:pa}
\end{figure} 

To understand whether this trend is due to variations in the viewing angle produced by jet bending  we compare the innermost jet position angle (PA) at 15~GHz from the MOJAVE survey\footnote{http://www.physics.purdue.edu/astro/MOJAVE/allsources.html} \citep{Pushkarev2012} and 43~GHz from the Boston University monitoring program\footnote{https://www.bu.edu/blazars/VLBAproject.html} \citep{Jorstad2017}. Figure  \ref{plt:pa} shows that comparison for the common sources in Sample A (22 sources) and Sample B (5 sources). Only 0420-014 from Sample B shows a discrepancy in the jet position angles suggesting the trend we find is due to a viewing angle change. A curved jet geometry for this source has already been noted in \cite{Britzen2000}. The remaining four sources from Sample B, namely 0716+714, 0954+658, 1253-055 and 1749+096, show similar position angles. Hence, the trend we detect in those sources, if real, is likely due to a velocity variation. Interestingly 0851+202 (also known as OJ~287) in Sample A shows a slightly different position angle between 15 and 43~GHz. However, the source is known to change its jet position angle on timescales of 1-2 years \citep{Cohen2017,Britzen2018}. Hence, this difference can be attributed to non-simultaneous PA measurements at the two frequencies.

\subsection{Transverse velocity structure}

 Recent observations of M87 revealed a transverse velocity structure \citep{Mertens2016}. A similar spine-sheath jet structure has been invoked to explain the discrepancy between $\rm \delta_\mathrm{var}$  implied by radio observations and spectral energy distribution modeling (e.g., \citealp{Ghisellini2005}). It is not unlikely that different frequencies not only probe different regions, but also a different underlying jet flow. In the standard spine-sheath model where the spine is characterized by a faster flow we would expect an increase of  $\rm \delta_\mathrm{var}$ towards higher frequencies which is only observed in 1253-055. \cite{Mertens2016} found a more complex slow-fast-slow configuration in M~87. In this case, higher frequencies could be dominated by the innermost slower flow with lower frequencies dominated by a faster flow, thus creating the observed trends.

\subsection{Parabolic versus conical geometry}
\begin{figure}
\centering
\resizebox{\hsize}{!} {\includegraphics[width=\hsize]{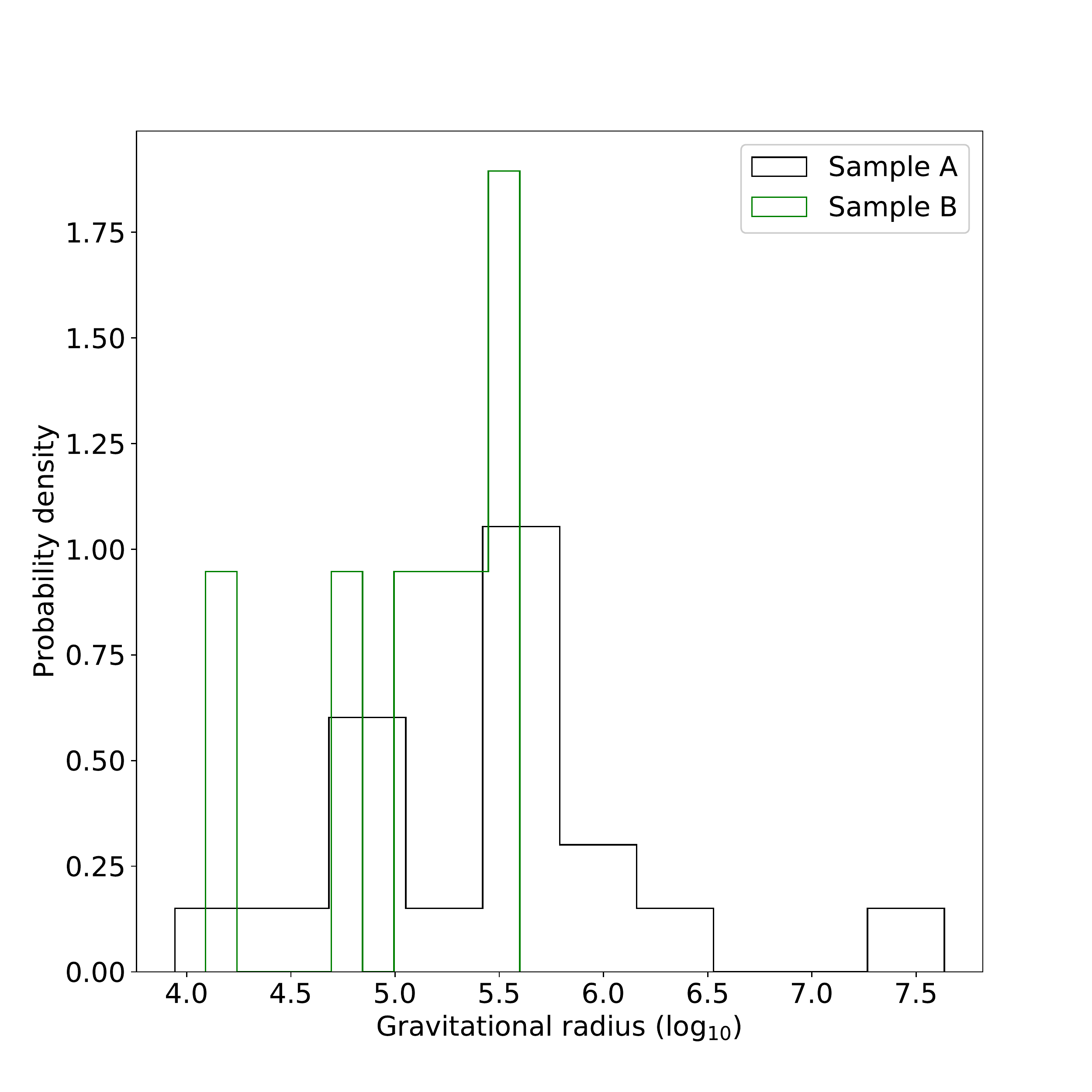}}
\caption{Distance of the 15~GHz core from the black hole in gravitational radii for Sample A (black) and Sample B (green). We do not find a statistically significant difference between the two samples (K-S test p-value=0.5)}
\label{plt:rg}
\end{figure} 

\begin{figure}
\centering
\resizebox{\hsize}{!} {\includegraphics[width=\hsize]{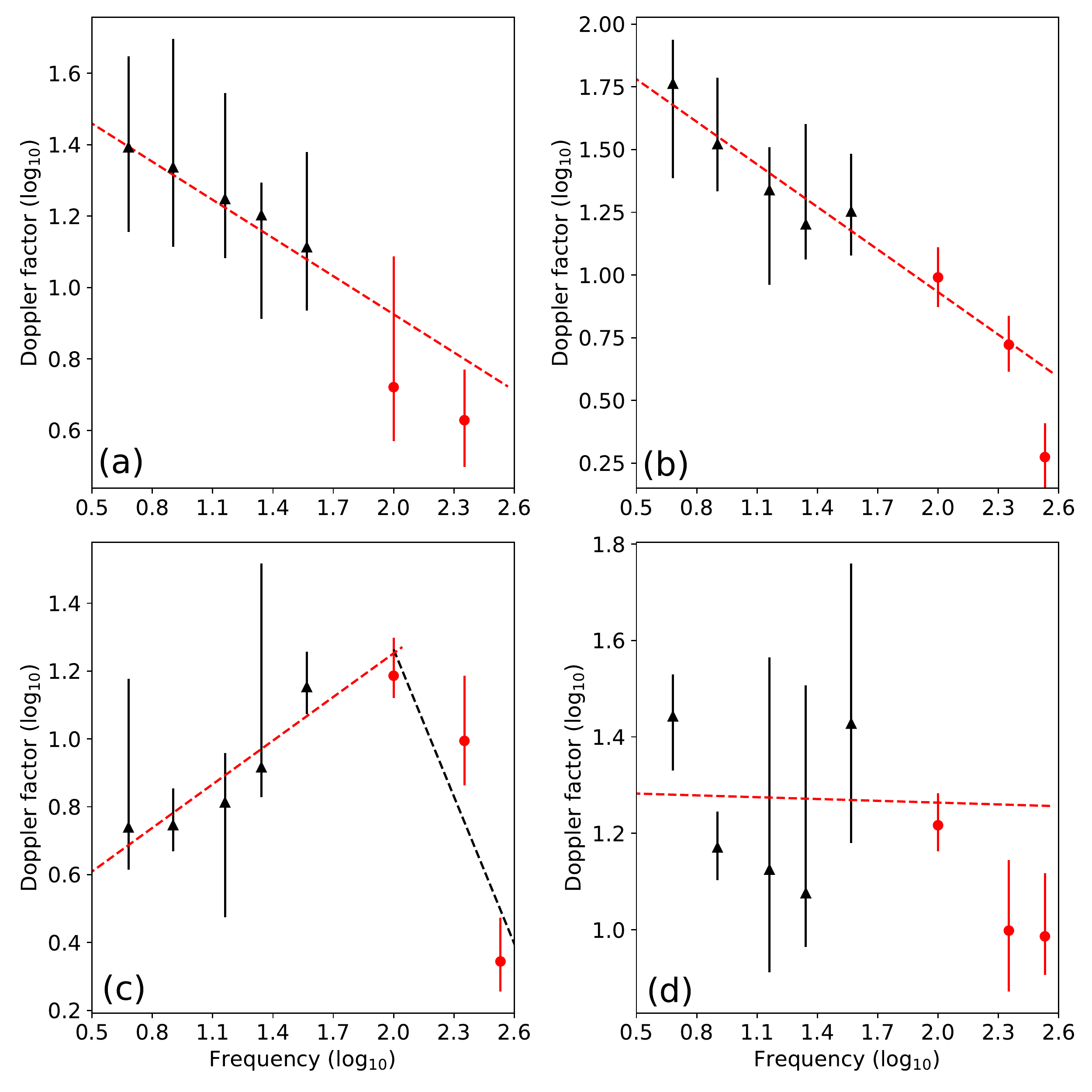}}
\caption{Doppler factor versus frequency in log-log space for 0234+285 (panel a) 0420-014 (panel b), 1253-055 (panel c) and 2251+158 (panel d). The red dashed line shows the best-fit relation estimated in the 4.8-37~GHz range.The dashed black line in panel c is the best-fit relation estimated in the 100-350~GHz range.}
\label{plt:highfreq}
\end{figure}

 The shape of the jet can be studied using VLBI observations and by determining the width of the jet $d$ as a function of distance $r$ from the radio core. This is usually modeled with a power-law function $d\propto{r^{k}}$, where $d$ can be estimated from a Gaussian fit to the transverse brightness profile with a FWHM $D$ and the FWHM of the restoring beam $b$ so that $d=(D^2-b^2)^{1/2}$ \citep{Pushkarev2017}. In a conical jet $k=1$, while in a parabolic jet $k=0.5$. According to the jet-transition model, the acceleration zone in blazars is expected to end at $\sim10^5 $ gravitational radii ($\rm R_g$, \citealp{Marscher2008,Asada2014}), where the jet is expected to change from parabolic to conical.

 We estimate the distance from the black hole to the 15~GHz core in $\rm R_g$ using the de-projected core distance estimates from \cite{Pushkarev2012} and black hole masses from \cite{Liodakis2020}. There are 18 sources in Sample A and 7 in Sample B with an available $\rm R_g$ estimate. Most of the sources cluster around $\rm \sim10^5~R_g$ (Fig. \ref{plt:rg}). One of the sources in Sample B has a distance $\rm <10^5~R_g$ (0212+735). Interestingly 0212+735 also shows a geometry at 15~GHz closer to parabolic (k-index=0.53). 0804+499 in Sample A shows a slightly lower value, although we do not detect a Doppler factor versus frequency trend. However, it is likely that the distance to the transition region ($\rm \sim10^5 R_g$) is not universal. The transition from a parabolic to a conical geometry can be different for different sources and occur closer to the black hole \citep{Boccardi2016,Boccardi2021}.  Different VLBI studies can also produce discrepant results depending on the time-span used in the analysis (e.g., \citealp{Boccardi2021,Park2021})). Interestingly, \cite{Boccardi2021} find a parabolic geometry (k-index $\leq0.6$) for four sources common with our Sample A, namely 0316+413, 0430+052, 0415+379, and 1807+698 (3 radio galaxies and 1 BL Lac object), where we do not detect a significant trend. This discrepancy could be related to either the caveats discussed below (section \ref{caveats}) or due to their low $\rm \delta_\mathrm{var}<5$ preventing us from identifying any trend.

To further test the jet-transition scenario we turn to high frequency ($>$37~GHz) observations. If the sources in Sample B are in the parabolic geometry regime we expect the trend we find in the $\rm \delta_\mathrm{var}$ versus frequency plane to continue towards higher frequencies consistent with the best-fit trend. On the other hand, if the sources in Sample A are in the conical regime, towards higher frequencies we expect  to find a transiting trend towards  lower values for $\rm \delta_\mathrm{var}$. We use data from the ALMA calibrator continuum observations catalog \citep{Bonato2018}\footnote{https://almascience.eso.org/sc/} and CARMA at $\sim$100~GHz (21 sources), and SMA \citep{Gurwell2007}\footnote{http://sma1.sma.hawaii.edu/callist/callist.html}\footnote{The SMA observations for a given source are taken at slightly different frequencies (a dispersion of typically a few GHz). Here we quote the median frequency for all sources. } at 225 (45 sources), and 340~GHz (14 sources) to estimate $\rm \delta_\mathrm{var}$ following the same procedure as above (section \ref{sampl}). There are a few additional sources from our sample included in those databases. However, those typically have less than 40 observations in total. We  therefore excluded them from our analysis. The earliest ALMA and CARMA observation is at MJD~55701 (May 2011) and the latest is at MJD~59292 (March 2021). For the SMA observations, the earliest is at MJD~52431 (June 2006) while the latest is at MJD~59267 (February 2021). Table \ref{table:3} lists the high-frequency $\rm \delta_\mathrm{var}$ estimates.

Overall, we find that the high-frequency estimates for Sample A tend to be lower than the best-fit trend. From Sample B, 0716+714, 0736+017, and 1749+096 show lower $\rm \delta_\mathrm{var}$ similar to Sample A sources. On the other hand, 0234+285 and 0420-014 from Sample B show the expected behavior (Fig. \ref{plt:highfreq} panels a, b). Given the difference in the innermost position angles found for 0420-014, this can then be interpreted as a continuously curved jet across the GHz range. Interestingly, 1253-055 shows a decreasing trend at high frequencies, although this is not statistically significant ( $\rho=$-0.88 p-value=0.3, Fig.  \ref{plt:highfreq} panel c). One interpretation could be that the jet is reaching terminal velocity close to 100~GHz and then decelerating. Recent Event Horizon telescope (EHT) observations at $\sim$230~GHz \citep{Kim2020} found that the jet is likely bent. This could explain the change of trends from the high to the low frequencies. Unlike 0420-014, the similar jet position angle between 43 and 15~GHz suggests that the trend of lower $\rm \delta_\mathrm{var}$ at lower frequencies is most likely due to deceleration. However, we note that the time window of the light curves used for the high-frequency modeling is shorter with little time overlap with the 4.8-37~GHz observations. Such a time difference can lead to the underestimation of the $\rm \delta_\mathrm{var}$ at the highest frequencies. Although tantalizing, our results for the higher frequencies should be treated with caution. 

Alternatively, acceleration at parsecs scales in a conical geometry can occur in a striped jet model with reversing toroidal magnetic field polarities \citep{Zhang2021}. In this case, jet acceleration is powered by magnetic energy dissipation via magnetic reconnection between stripes and can continue even after tens of parsecs from the black hole. This would suggest that Sample B sources likely host slower spinning supermassive black holes with a smaller stripe width spectral index ($\alpha$). We test this scenario using the spin estimates from \cite{Liodakis2018-III}. There are 32 sources from Sample A and 12 sources from Sample B with an available estimate. We find no statistically significant difference between the two samples (K-S p-value=0.97). We note that the spin estimates from \cite{Liodakis2018-III} are model dependent and might not be representative of the ``true'' black hole spin of the sources.

\subsection{Overall VLBI properties}

We additionally discuss below the VLBI properties of the two samples using data from the MOJAVE survey \citep{Kovalev2005,Pushkarev2012,Hovatta2014,Homan2015,Pushkarev2017,Hodge2018,Lister2019}. In the majority of cases we do not find a statistically significant difference, hence, we highlight only a few interesting comparisons.

We compare the median relative parallel and perpendicular acceleration  (to their proper motion vector, see Eqs. 5 \& 6 in \citealp{Homan2015}) of jet components in the two samples (28 sources from Sample A and 12 from Sample B). Parallel acceleration is often considered to reflect changes in the flow speed while perpendicular acceleration is believed to reflect changes in the direction. Starting from the relative parallel acceleration, we do not find a statistically significant difference (K-S p-value=0.33) between the two samples. If the measured acceleration of the jet components is representative of the jet bulk flow, one might expect Sample B sources to show higher acceleration.  This would likely suggest that either Sample B is a mixture of accelerating sources and sources with changing viewing angles or that the acceleration of jet components reflects velocity changes of hotspots moving within an underlying, quiescent flow (e.g., \citealp{Ghisellini2008}). Interestingly, Sample B sources show on average higher relative perpendicular acceleration (K-S test p-value=0.03). This would be in favor of either velocity stratification or jet bending producing the observed trends (see above). 

Sample B sources are on average more core dominated (K-S p-value=0.023, \citealp{Kovalev2005}). At the same time, the $\rm \delta_\mathrm{var}$ distributions at 15 GHz do not show a significant difference (K-S test p-value=0.56). This is in tension with the common assumption of the core dominance being a proxy for higher beaming, but our small sample size can also affect our conclusions. Sample A sources have on average higher maximum apparent jet velocity  ($\beta_{app,max}$, \citealp{Lister2019}). The K-S test rejects the null hypothesis when comparing the distributions for the two samples (p-value=0.024) whereas the A-D test does not reject it, albeit, marginally (p-value=0.056). Excluding 1253-055 (which shows the opposite trend to the rest of the Sample B sources), both tests reject the null hypothesis (p-value$<0.026$). 

Using $\beta_{app,max}$ and $\rm \delta_\mathrm{var,15}$ we can estimate the viewing angle and Lorentz factor distributions. We do not find a significant difference in the viewing angle distributions (K-S p-value=0.065). For the Lorentz factor distributions the Sample B sources have on average lower values according to the K-S test (p-value=0.02) which is not confirmed by the A-D test (p-value=0.062). However, this trend is confirmed by both tests (p-value$<$0.023) when excluding 1253-055. This could suggest that, on average, the remaining sources in Sample B (i.e., sources that show higher $\rm \delta_\mathrm{var}$ towards lower frequencies) have not yet reached Lorentz factors as high as those of Sample A sources at 15 GHz, i.e. they are still accelerating.

\subsection{Caveats}\label{caveats}

In the statistical analysis presented above, we used the available literature values. This often results in at least one sample (most often Sample B) having approximately or even fewer than 10 sources hampering the discriminating strength of the K-S and A-D tests. 

The cadence of the observations used for the light-curve modeling, if insufficient, can lead to an underestimation of the true Doppler factor in a given source \citep{Liodakis2015-II}. Sources in our sample have varying average cadence, from a few days to a few weeks. Combined with the fact that flare rise times are expected to be shorter at higher frequencies, this can lead to the underestimation of the 22 and 37~GHz Doppler factors creating an artificial trend. Out of Sample B, 0458-020, 1739+522, 1803+784, 2007+777, and 2121+053 have a factor of two lower cadence at both high frequencies; hence the results from those sources should be treated with caution. The majority of our sources have comparable sampling, bu it is nevertheless possible, although unlikely, that for some sources we are underestimating $\rm \delta_\mathrm{var}$ at the lower frequencies, hence destroying any intrinsic trend. If we are systematically underestimating the $\rm \delta_\mathrm{var}$ at 22 and 37~GHz, this would most likely suggest that the majority of our sources have decelerating jets or jets steering away from our line of sight.

Throughout this work, we have assumed the same value for the $T_\mathrm{int,max}$ for all frequencies. It could be possible for $T_\mathrm{int,max}$ to be different for different frequencies, if for example the balance between particle and magnetic field energy densities changes with distance from the core. However, recent VLBI results at 86~GHz suggest a $\rm T_\mathrm{int,max}\sim 3.7\times10^{11} K$ \citep{Nair2019}, consistent within the uncertainties from the 15~GHz value we used in this work. Hence, any $T_\mathrm{int,max}$ variations in the 4.8-37~GHz range are unlikely to have a significant impact on our results.

The $\rm \delta_\mathrm{var}$ estimates found in this work represent an on average $\rm \delta_\mathrm{var}$ for a given observing period. Individual flares, can nevertheless yield both higher and lower $\rm \delta_\mathrm{var}$. Changes in the viewing angle by factors of 2-3 during individual events have been noted in previous studies (e.g., \citealp{Larionov2010,Raiteri2017-II,Uemura2017,Liodakis2020-II}). Variations in other geometric and physical parameters of the emission region (e.g., Lorentz factor, magnetic field strength etc.) are also possible. This is likely imprinted in the shape of the flares \citep{Roy2019}, which we find to have both symmetric and asymmetric (either fast-rise-slow-decay or slow-rise-fast-decay) profiles. The aforementioned $\rm \delta_\mathrm{var}$ variations are reflected in the estimates' accompanying confidence intervals (Table \ref{table:1}, \ref{table:3}) which should not be treated as statistical, but instead as the possible range of  $\rm \delta_\mathrm{var}$ for a given source.

\section{Conclusions}\label{concl}

We studied the relativistic effects across five radio frequencies from 4.8 to 37~GHz for 61 sources. By quantifying the Doppler factor in each frequency we are able to study variations possibly related to acceleration, deceleration, or jet bending. The majority of the sources in our sample do not show any such variations across frequencies. This would be consistent with nonaccelerating conical jets. However, we identify 17 interesting sources; 16 show higher Doppler factor towards lower frequencies and one shows the opposite trend. To test the different possible origins of $\rm \delta_\mathrm{var}$ versus frequency trends we use the VLBI properties and high-frequency observations of the sources (100, 225, and 340~GHz, 45 sources have at least one high-frequency estimate) to estimate $\rm \delta_\mathrm{var}$ at such high frequencies for the first time. Our analysis suggests that the trend found in 0420-014 is likely due to jet bending, while the trend in 0212+735 is likely due to the jet accelerating in a parabolic geometry. 1253-055 shows a complex behavior that is likely attributed to a bend in the innermost jet probed by the highest frequencies, while decelerating at the 4.8-37~GHz range. For the remaining sources, our results are broadly consistent with the expectations from a transitioning geometry model with a few exceptions. However, the much shorter time-span of the high-frequency observations prevents us from coming to strong conclusions. Simultaneous high-cadence monitoring across the entire GHz-millimeter range, which will be available in the future with the Simons observatory \citep{SimonsObservatory2019} and CMB S4 \citep{CMBS42016}, will provide an unprecedented opportunity to study the structure of blazar jets through their multiwavelength radio variability.

\begin{acknowledgements}
The authors would like to thank Carolina Casadio, Dan Homan, and the anonymous referee for useful comments and discussions that helped improve this work. I. L. thanks the University of Crete for their hospitality during which the paper was written. T. H. was supported by the Academy of Finland projects 317383, 320085, and 322535. This research has made use of data from the MOJAVE database that is maintained by the MOJAVE team  \citep{Lister2018}. This study makes use of 43 GHz VLBA data from the VLBA-BU Blazar Monitoring Program (VLBA-BU-BLAZAR;
http://www.bu.edu/blazars/VLBAproject.html), funded by NASA through the Fermi Guest Investigator Program. The VLBA is an instrument of the National Radio Astronomy Observatory. The National Radio Astronomy Observatory is a facility of the National Science Foundation operated by Associated Universities, Inc. This research has made use of data from the University of Michigan Radio Astronomy Observatory which has been supported by the University of Michigan and by a series of grants from the National Science Foundation, most recently AST-0607523, and from NASA Fermi G. I. grants NNX09AU16G, NNX10AP16G, NNX13AP18G, and NNX11AO13G. This publication makes use of data obtained at Mets\"ahovi Radio Observatory, operated by Aalto University in Finland. The Submillimeter Array is a joint project between the Smithsonian Astrophysical Observatory and the Academia Sinica Institute of Astronomy and Astrophysics and is funded by the Smithsonian Institution and the Academia Sinica.
\end{acknowledgements}

%
  \bibliographystyle{aa} 
  \bibliography{bibliography.bib} 

\begin{appendix}
\section{Tables.}
\onecolumn
\renewcommand{\arraystretch}{1.5}
\begin{longtable}{ l c c c c c c} 

\caption{Median Doppler factor estimates and their 68\% confidence intervals for all the sources in our sample. Column (1) is the B1950 name, column (2) is the redshift, column (3) is $\delta_{var}$ at 4.8~GHz,  column (4) is $\delta_{var}$ at 8~GHz, column (5) is $\delta_{var}$ at 14.5~GHz, column (6) is $\delta_{var}$ at 22~GHz, and  column (7) is $\delta_{var}$ at 37~GHz.} \label{table:1}   \\         
\hline\hline                 
Name & z & $\delta_{var,4.8}$ & $\delta_{var,8}$ & $\delta_{var,14.5}$ &
$\delta_{var,22}$ & $\delta_{var,37}$ \\   
\hline                        
0048-097 & 0.635 & $13.33^{+11.79}_{-10.67}$ & $14.34^{+12.36}_{-6.64}$ & $8.35^{+7.71}_{-4.37}$ & $4.04^{+1.79}_{-1.84}$ & - \\ 
0059+581 & 0.644 & $49.18^{+8.94}_{-10.91}$ & $29.65^{+8.85}_{-10.34}$ & $23.05^{+11.13}_{-8.44}$ & $12.69^{+5.76}_{-3.68}$ & $8.34^{+4.77}_{-2.24}$ \\ 
0106+013 & 2.107 & $20.06^{+30.99}_{-7.73}$ & $22.46^{+16.94}_{-13.14}$ & $19.22^{+8.86}_{-5.14}$ & $14.59^{+5.91}_{-7.96}$ & $13.36^{+3.71}_{-6.81}$ \\ 
0109+224 & 0.265 & $4.55^{+3.17}_{-0.99}$ & $6.24^{+1.62}_{-1.63}$ & $5.51^{+1.8}_{-1.39}$ & $4.73^{+1.77}_{-1.52}$ & $5.5^{+1.46}_{-1.52}$ \\ 
0133+476 & 0.859 & $29.71^{+12.8}_{-6.87}$ & $9.14^{+17.61}_{-7.14}$ & $16.47^{+6.07}_{-14.31}$ & $16.96^{+3.75}_{-3.27}$ & $18.99^{+4.33}_{-5.18}$ \\ 
0202+149 & 0.834 & $14.9^{+10.38}_{-7.0}$ & $5.36^{+5.18}_{-2.91}$ & $12.19^{+9.56}_{-6.39}$ & $6.99^{+5.98}_{-2.71}$ & $8.45^{+1.86}_{-2.41}$ \\ 
0212+735 & 2.367 & - & $40.95^{+18.72}_{-9.64}$ & $25.73^{+12.51}_{-7.37}$ & $22.78^{+7.01}_{-9.46}$ & $10.81^{+4.26}_{-4.01}$ \\ 
0234+285 & 1.213 & $24.7^{+14.51}_{-13.51}$ & $21.71^{+18.0}_{-11.15}$ & $17.7^{+12.11}_{-6.76}$ & $15.94^{+3.36}_{-10.68}$ & $12.98^{+7.95}_{-5.31}$ \\ 
0235+164 & 0.94 & $53.71^{+15.63}_{-30.67}$ & $34.77^{+22.9}_{-16.79}$ & $26.46^{+10.2}_{-14.77}$ & $27.24^{+6.73}_{-6.43}$ & $27.54^{+18.6}_{-16.21}$ \\ 
0300+470 & 0.475 & $8.91^{+9.95}_{-7.16}$ & $4.41^{+7.86}_{-1.57}$ & $7.84^{+5.16}_{-2.23}$ & $2.24^{+1.74}_{-1.71}$ & $1.71^{+1.18}_{-0.86}$ \\ 
0306+102 & 0.863 & $8.23^{+2.17}_{-4.54}$ & $7.69^{+3.89}_{-2.11}$ & $9.54^{+1.45}_{-2.73}$ & $4.37^{+1.51}_{-1.69}$ & $2.75^{+1.79}_{-1.51}$ \\ 
0316+413 & 0.018 & $0.2^{+0.06}_{-0.05}$ & $0.22^{+0.08}_{-0.06}$ & $0.19^{+0.37}_{-0.07}$ & $0.2^{+1.67}_{-0.09}$ & $0.13^{+1.5}_{-0.03}$ \\
0336-019 & 0.852 & $22.82^{+6.46}_{-10.01}$ & $16.28^{+4.06}_{-3.08}$ & $18.99^{+5.97}_{-3.86}$ & $7.85^{+3.19}_{-2.73}$ & $13.79^{+7.23}_{-4.82}$ \\ 
0415+379 & 0.048 & $1.92^{+0.69}_{-0.21}$ & $2.94^{+0.71}_{-1.39}$ & $2.46^{+3.55}_{-1.24}$ & $2.36^{+0.5}_{-0.46}$ & $3.48^{+0.83}_{-0.85}$ \\ 
0420-014 & 0.915 & $57.92^{+23.21}_{-50.24}$ & $33.26^{+20.22}_{-14.39}$ & $21.8^{+8.54}_{-18.92}$ & $15.92^{+14.68}_{-5.12}$ & $17.9^{+9.46}_{-7.26}$ \\ 
0422+004 & 0.268 & $7.25^{+4.92}_{-5.77}$ & $6.98^{+3.72}_{-2.52}$ & $5.14^{+4.26}_{-1.66}$ & $4.04^{+1.02}_{-1.01}$ & $3.79^{+1.42}_{-1.25}$ \\ 
0430+052 & 0.033 & $2.82^{+2.07}_{-0.63}$ & $1.81^{+1.98}_{-0.58}$ & $2.59^{+0.83}_{-1.58}$ & $2.8^{+0.78}_{-0.55}$ & $2.48^{+0.77}_{-1.21}$ \\ 
0458-020 & 2.291 & $37.66^{+32.61}_{-24.62}$ & $12.66^{+17.6}_{-7.13}$ & $8.52^{+15.31}_{-1.3}$ & $4.57^{+11.32}_{-3.19}$ & $3.36^{+3.95}_{-2.43}$ \\ 
0528+134 & 2.07 & $39.98^{+15.75}_{-24.35}$ & $40.23^{+20.68}_{-34.92}$ & $35.92^{+31.81}_{-20.32}$ & $47.26^{+33.51}_{-26.47}$ & $28.01^{+21.12}_{-22.53}$ \\ 
0605-085 & 0.872 & $2.53^{+0.3}_{-0.18}$ & $6.65^{+3.17}_{-3.45}$ & $2.71^{+4.97}_{-0.35}$ & $2.06^{+1.5}_{-1.01}$ & $2.89^{+2.19}_{-2.21}$ \\ 
0716+714 & 0.31 & $19.94^{+11.99}_{-9.87}$ & $16.71^{+6.51}_{-6.93}$ & $13.64^{+4.46}_{-4.67}$ & $13.83^{+4.55}_{-4.07}$ & $12.41^{+5.4}_{-3.08}$ \\ 
0735+178 & 0.424 & $2.39^{+8.01}_{-1.59}$ & $6.18^{+6.13}_{-3.27}$ & $5.93^{+8.82}_{-1.68}$ & $5.31^{+4.46}_{-1.8}$ & $3.93^{+4.55}_{-2.14}$ \\ 
0736+017 & 0.191 & $7.61^{+2.34}_{-2.19}$ & $7.8^{+2.99}_{-2.06}$ & $6.19^{+2.8}_{-1.99}$ & $5.57^{+1.22}_{-1.44}$ & $5.63^{+1.37}_{-1.42}$ \\ 
0754+100 & 0.266 & $4.47^{+1.57}_{-1.19}$ & $7.43^{+1.04}_{-1.62}$ & $4.8^{+3.11}_{-1.54}$ & $4.84^{+1.9}_{-2.16}$ & $3.06^{+0.89}_{-1.14}$ \\ 
0804+499 & 1.436 & $36.61^{+17.73}_{-5.87}$ & $12.55^{+14.92}_{-5.05}$ & $30.49^{+12.93}_{-11.48}$ & $28.4^{+8.18}_{-8.66}$ & $30.84^{+7.87}_{-14.29}$ \\ 
0814+425 & 1.381 & $19.34^{+10.08}_{-6.19}$ & $31.02^{+25.09}_{-22.33}$ & $9.43^{+4.3}_{-3.4}$ & $13.57^{+5.5}_{-8.8}$ & $7.05^{+1.79}_{-2.54}$ \\ 
0851+202 & 0.306 & $15.17^{+9.57}_{-5.38}$ & $25.16^{+8.26}_{-9.41}$ & $17.04^{+7.37}_{-7.87}$ & $19.46^{+15.42}_{-4.64}$ & $27.97^{+13.29}_{-10.45}$ \\ 
0954+658 & 0.367 & $21.61^{+7.5}_{-10.87}$ & $14.17^{+3.53}_{-3.47}$ & $6.65^{+4.42}_{-1.72}$ & $4.46^{+1.18}_{-2.05}$ & $5.87^{+3.57}_{-1.92}$ \\ 
1055+018 & 0.888 & $12.78^{+11.28}_{-7.57}$ & $12.75^{+6.59}_{-9.83}$ & $9.49^{+17.25}_{-5.09}$ & $15.4^{+5.54}_{-3.33}$ & $18.05^{+7.58}_{-9.18}$ \\ 
1101+3828 & 0.03 & $3.33^{+1.01}_{-0.85}$ & $3.19^{+0.74}_{-0.92}$ & $2.25^{+0.5}_{-0.73}$ & $2.81^{+0.53}_{-1.0}$ & $2.55^{+1.44}_{-1.39}$ \\
1156+295 & 0.729 & $19.04^{+8.94}_{-15.03}$ & $32.82^{+13.04}_{-22.63}$ & $22.09^{+3.96}_{-15.58}$ & $15.62^{+4.83}_{-3.6}$ & $33.32^{+18.88}_{-20.58}$ \\ 
1219+285 & 0.102 & $0.86^{+0.6}_{-0.12}$ & $0.6^{+1.52}_{-0.1}$ & $1.45^{+0.81}_{-0.63}$ & $0.47^{+0.9}_{-0.27}$ & $0.47^{+1.24}_{-0.36}$ \\ 
1222+216 & 0.435 & $6.7^{+3.0}_{-3.42}$ & $3.48^{+8.36}_{-2.83}$ & $4.13^{+3.4}_{-2.81}$ & $3.31^{+5.91}_{-1.31}$ & $3.43^{+2.55}_{-0.87}$ \\ 
1226+023 & 0.158 & $4.27^{+2.43}_{-3.12}$ & $5.71^{+4.34}_{-4.22}$ & $4.43^{+1.04}_{-0.85}$ & $6.0^{+2.37}_{-1.16}$ & $4.25^{+3.09}_{-1.26}$ \\ 
1253-055 & 0.536 & $5.49^{+5.54}_{-1.58}$ & $5.57^{+1.38}_{-0.99}$ & $6.5^{+2.17}_{-5.07}$ & $8.25^{+11.4}_{-1.68}$ & $14.23^{+3.38}_{-2.61}$ \\
1308+326 & 0.997 & $10.4^{+10.89}_{-8.21}$ & $13.38^{+14.23}_{-7.86}$ & $9.19^{+3.87}_{-5.63}$ & $13.94^{+7.87}_{-4.09}$ & $13.87^{+7.11}_{-3.72}$ \\ 
1335-127 & 0.539 & $9.57^{+8.34}_{-2.14}$ & $13.27^{+10.27}_{-5.37}$ & $10.29^{+4.03}_{-3.47}$ & $6.62^{+10.86}_{-2.45}$ & $5.95^{+4.49}_{-1.47}$ \\ 
1413+135 & 0.247 & $4.05^{+2.49}_{-2.34}$ & $5.36^{+8.56}_{-1.83}$ & $7.41^{+2.62}_{-4.05}$ & $4.67^{+1.64}_{-2.26}$ & $2.63^{+0.95}_{-0.69}$ \\ 
1418+546 & 0.152 & $7.77^{+5.57}_{-5.55}$ & $5.9^{+1.66}_{-1.77}$ & $3.72^{+0.76}_{-2.34}$ & $2.78^{+1.33}_{-0.9}$ & $2.55^{+1.38}_{-2.32}$ \\ 
1502+106 & 1.839 & $9.97^{+9.59}_{-8.58}$ & $25.87^{+26.44}_{-11.26}$ & $31.21^{+13.37}_{-10.2}$ & $10.02^{+6.4}_{-4.58}$ & $12.27^{+8.19}_{-4.08}$ \\ 
1510-089 & 0.36 & $24.7^{+9.76}_{-7.77}$ & $23.04^{+11.43}_{-10.97}$ & $22.94^{+10.5}_{-11.91}$ & $20.58^{+7.76}_{-5.67}$ & $24.02^{+6.46}_{-11.13}$ \\ 
1553+113 & 0.36 & $1.56^{+3.57}_{-1.49}$ & $0.34^{+2.68}_{-0.26}$ & $0.68^{+3.68}_{-0.47}$ & $1.13^{+1.05}_{-0.44}$ & - \\ 
1611+343 & 1.397 & $4.64^{+2.37}_{-1.42}$ & $9.25^{+8.47}_{-6.41}$ & $11.12^{+12.87}_{-7.59}$ & $7.4^{+5.55}_{-4.51}$ & $9.54^{+9.47}_{-5.72}$ \\ 
1633+382 & 1.814 & $29.02^{+28.37}_{-12.2}$ & $37.89^{+15.03}_{-14.1}$ & $39.07^{+10.16}_{-13.49}$ & $24.99^{+15.97}_{-10.22}$ & $45.15^{+23.92}_{-13.98}$ \\ 
1641+399 & 0.593 & $6.13^{+23.99}_{-1.61}$ & $5.93^{+1.19}_{-2.65}$ & $3.23^{+5.54}_{-1.05}$ & $4.58^{+1.61}_{-0.84}$ & $7.58^{+3.79}_{-3.67}$ \\ 
1642+690 & 0.751 & $5.24^{+4.94}_{-1.76}$ & $10.15^{+3.53}_{-4.41}$ & $4.57^{+2.71}_{-2.24}$ & $3.58^{+2.87}_{-2.15}$ & $3.78^{+3.26}_{-2.57}$ \\ 
1730-130 & 0.902 & $55.67^{+21.55}_{-24.35}$ & $12.22^{+22.06}_{-2.04}$ & $9.7^{+13.02}_{-4.79}$ & $4.43^{+4.58}_{-1.75}$ & $9.5^{+4.71}_{-4.38}$ \\ 
1739+522 & 1.379 & $51.9^{+16.1}_{-13.21}$ & $24.68^{+13.6}_{-9.64}$ & $27.13^{+10.51}_{-8.27}$ & $12.95^{+5.87}_{-2.5}$ & $8.28^{+2.46}_{-3.29}$ \\ 
1741-038 & 1.057 & $20.21^{+22.15}_{-7.72}$ & $18.46^{+13.75}_{-8.35}$ & $22.76^{+21.51}_{-15.15}$ & $15.38^{+9.33}_{-4.32}$ & $23.11^{+7.74}_{-8.29}$ \\ 
1749+096 & 0.322 & $19.65^{+13.72}_{-8.73}$ & $18.75^{+4.57}_{-5.16}$ & $17.06^{+6.67}_{-6.07}$ & $17.22^{+7.06}_{-7.11}$ & $13.98^{+6.16}_{-2.29}$ \\ 
1803+784 & 0.68 & $26.94^{+10.86}_{-10.39}$ & $19.35^{+15.23}_{-9.68}$ & $19.24^{+5.59}_{-6.0}$ & $6.81^{+2.54}_{-1.77}$ & $8.33^{+2.08}_{-2.04}$ \\ 
1807+698 & 0.051 & $2.34^{+0.73}_{-0.69}$ & $4.2^{+2.09}_{-2.94}$ & $1.08^{+0.93}_{-0.81}$ & $1.21^{+0.47}_{-0.46}$ & $0.87^{+0.34}_{-0.21}$ \\ 
1823+568 & 0.664 & $8.23^{+6.33}_{-4.8}$ & $9.4^{+10.13}_{-6.87}$ & $5.23^{+3.07}_{-2.57}$ & $2.88^{+1.79}_{-1.96}$ & $3.67^{+1.86}_{-1.82}$ \\ 
2005+403 & 1.736 & $21.66^{+7.94}_{-6.97}$ & $34.51^{+22.24}_{-11.45}$ & $14.8^{+18.52}_{-14.01}$ & $21.08^{+6.56}_{-7.71}$ & $12.75^{+8.01}_{-11.35}$ \\ 
2007+777 & 0.342 & $12.39^{+3.47}_{-3.9}$ & $13.57^{+7.32}_{-4.34}$ & $6.22^{+6.02}_{-3.14}$ & $4.23^{+1.41}_{-1.34}$ & $2.94^{+1.48}_{-1.18}$ \\ 
2121+053 & 1.941 & $48.04^{+15.49}_{-18.62}$ & $47.09^{+20.05}_{-10.58}$ & $27.75^{+16.61}_{-14.94}$ & $11.34^{+6.01}_{-4.11}$ & $14.32^{+5.46}_{-6.25}$ \\ 
2200+420 & 0.069 & $8.43^{+6.46}_{-6.68}$ & $9.75^{+3.43}_{-3.34}$ & $7.93^{+3.91}_{-3.58}$ & $7.34^{+3.51}_{-1.93}$ & $9.95^{+5.06}_{-4.39}$ \\ 
2223-052 & 1.404 & $10.08^{+29.59}_{-5.47}$ & $37.38^{+50.69}_{-24.05}$ & $21.0^{+20.54}_{-15.18}$ & $13.38^{+6.12}_{-6.08}$ & $21.68^{+7.95}_{-8.11}$ \\ 
2230+114 & 1.037 & $31.82^{+26.39}_{-25.91}$ & $23.45^{+14.49}_{-13.6}$ & $18.41^{+28.4}_{-10.33}$ & $26.51^{+11.7}_{-9.55}$ & $22.65^{+11.11}_{-10.21}$ \\ 
2251+158 & 0.859 & $27.72^{+5.53}_{-7.15}$ & $14.82^{+2.53}_{-2.3}$ & $13.33^{+13.5}_{-6.55}$ & $11.91^{+11.83}_{-3.05}$ & $26.78^{+20.46}_{-15.26}$ \\ 
2254+074 & 0.19 & $2.14^{+2.22}_{-1.06}$ & $2.65^{+1.54}_{-1.27}$ & $1.52^{+0.46}_{-0.29}$ & $1.52^{+0.96}_{-1.08}$ & $1.11^{+0.58}_{-0.6}$ \\   
\hline                                   
\end{longtable}
\twocolumn

\onecolumn
\renewcommand{\arraystretch}{1.5}
\begin{longtable}{l c c c c c}

\caption{Doppler factor versus frequency correlation results. Column (1) is the B1950 name, column (2) is the median Pearson correlation $\rho$,  column (3) is the median Pearson correlation p-value, column (4) is the median best-fit slope, column (5) is the median best-fit intercept, and column (6) is the sample designation. ``A'' is for sources that do not show a statistically significant trend, ``B'' is for sources that do.\label{table:2}   }\\             
\hline\hline                 
Name & $\rho$ & p-value & slope & intercept & Sample\\    
\hline                        
0048-097  & -0.9 & 0.0998 & $-0.85^{+0.66}_{-0.48}$ & $1.84^{+0.49}_{-0.75}$ & A \\ 
0059+581  & -0.99 & 0.0014 & $-0.83^{+0.21}_{-0.19}$ & $2.25^{+0.2}_{-0.24}$ & B \\ 
0106+013  & -0.88 & 0.0504 & $-0.38^{+0.36}_{-0.39}$ & $1.68^{+0.48}_{-0.46}$ & A \\ 
0109+224  & 0.14 & 0.8271 & $-0.03^{+0.19}_{-0.22}$ & $0.76^{+0.26}_{-0.24}$ & A \\ 
0133+476  & -0.11 & 0.8654 & $-0.17^{+0.26}_{-0.23}$ & $1.44^{+0.29}_{-0.38}$ & A \\ 
0202+149  & -0.31 & 0.6098 & $-0.17^{+0.34}_{-0.3}$ & $1.15^{+0.38}_{-0.45}$ & A \\ 
0212+735  & -0.97 & 0.0313 & $-0.87^{+0.3}_{-0.36}$ & $2.44^{+0.43}_{-0.38}$ & B \\ 
0234+285  & -1.0 & 0.0002 & $-0.36^{+0.38}_{-0.31}$ & $1.64^{+0.37}_{-0.44}$ & B \\ 
0235+164  & -0.85 & 0.0664 & $-0.27^{+0.36}_{-0.32}$ & $1.8^{+0.37}_{-0.46}$ & A \\ 
0300+470  & -0.83 & 0.0809 & $-0.87^{+0.48}_{-0.42}$ & $1.63^{+0.53}_{-0.55}$ & A \\ 
0306+102  & -0.81 & 0.0951 & $-0.5^{+0.31}_{-0.32}$ & $1.33^{+0.32}_{-0.39}$ & A \\ 
0316+413  & -0.77 & 0.1303 & $0.48^{+0.61}_{-0.6}$ & $-1.07^{+0.55}_{-0.53}$ & A \\ 
0336-019  & -0.65 & 0.2305 & $-0.3^{+0.25}_{-0.24}$ & $1.53^{+0.25}_{-0.31}$ & A \\ 
0415+379  & 0.67 & 0.2114 & $0.19^{+0.17}_{-0.17}$ & $0.22^{+0.2}_{-0.23}$ & A \\ 
0420-014  & -0.93 & 0.0213 & $-0.57^{+0.37}_{-0.29}$ & $2.06^{+0.32}_{-0.47}$ & B \\ 
0422+004  & -0.97 & 0.0058 & $-0.36^{+0.37}_{-0.28}$ & $1.15^{+0.33}_{-0.51}$ & B \\ 
0430+052  & 0.16 & 0.7962 & $-0.08^{+0.26}_{-0.29}$ & $0.49^{+0.33}_{-0.33}$ & A \\ 
0458-020  & -0.98 & 0.0044 & $-1.06^{+0.49}_{-0.49}$ & $2.26^{+0.54}_{-0.58}$ & B \\ 
0528+134  & -0.46 & 0.4346 & $-0.04^{+0.42}_{-0.44}$ & $1.65^{+0.43}_{-0.56}$ & A \\ 
0605-085  & -0.32 & 0.5977 & $-0.16^{+0.29}_{-0.36}$ & $0.7^{+0.34}_{-0.3}$ & A \\ 
0716+714  & -0.96 & 0.0092 & $-0.2^{+0.3}_{-0.24}$ & $1.38^{+0.31}_{-0.42}$ & B \\ 
0735+178  & 0.35 & 0.5591 & $0.0^{+0.56}_{-0.52}$ & $0.75^{+0.65}_{-0.73}$ & A \\ 
0736+017  & -0.92 & 0.0258 & $-0.19^{+0.19}_{-0.17}$ & $1.03^{+0.21}_{-0.23}$ & B \\ 
0754+100  & -0.59 & 0.2914 & $-0.25^{+0.19}_{-0.25}$ & $0.95^{+0.24}_{-0.22}$ & A \\ 
0804+499  & -0.84 & 0.0733 & $-0.17^{+0.19}_{-0.26}$ & $1.69^{+0.25}_{-0.21}$ & A \\ 
0814+425  & -0.79 & 0.1134 & $-0.64^{+0.29}_{-0.26}$ & $1.88^{+0.28}_{-0.4}$ & A \\ 
0851+202  & 0.58 & 0.3099 & $0.22^{+0.29}_{-0.29}$ & $1.07^{+0.32}_{-0.35}$ & A \\ 
0954+658  & -0.91 & 0.0313 & $-0.72^{+0.28}_{-0.25}$ & $1.75^{+0.28}_{-0.33}$ & B \\ 
1055+018  & 0.54 & 0.3491 & $0.16^{+0.37}_{-0.35}$ & $0.96^{+0.41}_{-0.5}$ & A \\ 
1101+384  & -0.68 & 0.2022 & $-0.15^{+0.24}_{-0.34}$ & $0.59^{+0.32}_{-0.28}$ & A \\ 
1156+295  & 0.18 & 0.7758 & $0.1^{+0.43}_{-0.36}$ & $1.23^{+0.39}_{-0.57}$ & A \\ 
1219+285  & -0.44 & 0.4578 & $-0.25^{+0.45}_{-0.45}$ & $0.21^{+0.45}_{-0.45}$ & A \\ 
1222+216  & -0.75 & 0.1479 & $-0.19^{+0.37}_{-0.36}$ & $0.9^{+0.43}_{-0.52}$ & A \\ 
1226+023  & 0.02 & 0.9727 & $0.08^{+0.41}_{-0.31}$ & $0.62^{+0.35}_{-0.53}$ & A \\ 
1253-055 & 0.91 & 0.0322 & $0.43^{+0.23}_{-0.27}$ & $0.4^{+0.36}_{-0.27}$ & B \\ 
1308+326  & 0.47 & 0.4237 & $0.06^{+0.45}_{-0.35}$ & $1.01^{+0.44}_{-0.6}$ & A \\ 
1335-127  & -0.78 & 0.1236 & $-0.3^{+0.29}_{-0.3}$ & $1.33^{+0.35}_{-0.33}$ & A \\ 
1413+135  & -0.38 & 0.5244 & $-0.26^{+0.33}_{-0.3}$ & $0.94^{+0.39}_{-0.41}$ & A \\ 
1418+546  & -0.98 & 0.0026 & $-0.62^{+0.35}_{-0.4}$ & $1.3^{+0.41}_{-0.44}$ & B \\ 
1502+106  & -0.12 & 0.8482 & $-0.14^{+0.48}_{-0.37}$ & $1.39^{+0.42}_{-0.62}$ & A \\ 
1510-089  & -0.38 & 0.528 & $-0.08^{+0.25}_{-0.25}$ & $1.44^{+0.28}_{-0.33}$ & A \\ 
1553+113  & -0.07 & 0.9331 & $-0.2^{+0.82}_{-0.8}$ & $0.35^{+0.85}_{-0.96}$ & A \\ 
1611+343  & 0.59 & 0.2927 & $0.26^{+0.36}_{-0.39}$ & $0.66^{+0.4}_{-0.45}$ & A \\ 
1633+382  & 0.33 & 0.5925 & $0.04^{+0.34}_{-0.31}$ & $1.49^{+0.37}_{-0.4}$ & A \\ 
1641+399  & 0.04 & 0.9452 & $-0.25^{+0.47}_{-0.47}$ & $1.08^{+0.61}_{-0.6}$ & A \\ 
1642+690  & -0.65 & 0.2395 & $-0.36^{+0.35}_{-0.43}$ & $1.12^{+0.46}_{-0.4}$ & A \\ 
1730-130  & -0.78 & 0.1202 & $-0.93^{+0.35}_{-0.34}$ & $2.2^{+0.38}_{-0.41}$ & A \\ 
1739+522  & -0.95 & 0.0127 & $-0.86^{+0.23}_{-0.24}$ & $2.29^{+0.26}_{-0.29}$ & B \\ 
1741-038  & 0.11 & 0.864 & $-0.04^{+0.32}_{-0.33}$ & $1.36^{+0.39}_{-0.43}$ & A \\ 
1749+096  & -0.93 & 0.0222 & $-0.13^{+0.28}_{-0.28}$ & $1.4^{+0.32}_{-0.37}$ & B \\ 
1803+784  & -0.88 & 0.0473 & $-0.66^{+0.24}_{-0.23}$ & $1.9^{+0.27}_{-0.32}$ & B \\ 
1807+698  & -0.8 & 0.1023 & $-0.62^{+0.25}_{-0.22}$ & $0.9^{+0.28}_{-0.33}$ & A \\ 
1823+568  & -0.87 & 0.0549 & $-0.61^{+0.44}_{-0.41}$ & $1.41^{+0.46}_{-0.54}$ & A \\ 
2005+403  & -0.65 & 0.2305 & $-0.31^{+0.26}_{-0.37}$ & $1.68^{+0.35}_{-0.35}$ & A \\ 
2007+777  & -0.96 & 0.0089 & $-0.79^{+0.23}_{-0.28}$ & $1.73^{+0.28}_{-0.29}$ & B \\ 
2121+053  & -0.91 & 0.0339 & $-0.75^{+0.26}_{-0.27}$ & $2.25^{+0.28}_{-0.32}$ & B \\ 
2200+420  & 0.04 & 0.9439 & $-0.01^{+0.41}_{-0.32}$ & $0.96^{+0.35}_{-0.55}$ & A \\ 
2223-052  & 0.17 & 0.7883 & $-0.17^{+0.51}_{-0.46}$ & $1.51^{+0.61}_{-0.67}$ & A \\ 
2230+114  & -0.47 & 0.4209 & $-0.15^{+0.43}_{-0.35}$ & $1.54^{+0.41}_{-0.55}$ & A \\ 
2251+158  & -0.13 & 0.8358 & $-0.01^{+0.25}_{-0.3}$ & $1.29^{+0.28}_{-0.25}$ & A \\ 
2254+074  & -0.88 & 0.0471 & $-0.43^{+0.37}_{-0.41}$ & $0.74^{+0.45}_{-0.48}$ & B \\ 

\hline                                   
\end{longtable}

\twocolumn

\onecolumn
\renewcommand{\arraystretch}{1.5}
\begin{longtable}{l c c c }
\caption{High-frequency Doppler factor estimates for the sources in our sample. Column (1) is the B1950 name, column (2) is $\delta_{var}$ at 100~GHz,  column (3) is $\delta_{var}$ at 225~GHz, column (4) is $\delta_{var}$ at 340~GHz.\label{table:3} }\\
\hline\hline               
Name & $\delta_{var,100}$ & $\delta_{var,225}$ & $\delta_{var,340}$  \\    
\hline                        
0048-097 & - & $0.47^{+0.2}_{-0.13}$ & - \\ 
0059+581 & - & $4.11^{+1.27}_{-0.69}$ & - \\ 
0133+476 & - & $3.78^{+0.72}_{-0.82}$ & - \\ 
0234+285 & $5.26^{+4.44}_{-1.83}$ & $4.25^{+1.39}_{-1.28}$ & - \\ 
0235+164 & $6.84^{+2.81}_{-2.43}$ & $5.96^{+1.28}_{-1.41}$ & $2.23^{+0.61}_{-0.54}$ \\ 
0300+470 & - & $1.36^{+0.62}_{-0.39}$ & - \\ 
0306+102 & - & $1.97^{+0.78}_{-0.54}$ & - \\ 
0316+413 & $0.59^{+0.25}_{-0.28}$ & $0.28^{+0.14}_{-0.12}$ & $0.08^{+0.05}_{-0.06}$ \\
0336-019 & - & $2.6^{+0.55}_{-0.4}$ & - \\ 
0415+379 & - & $1.57^{+0.26}_{-0.29}$ & $0.29^{+0.08}_{-0.09}$ \\ 
0420-014 & $9.78^{+2.69}_{-2.67}$ & $5.27^{+1.4}_{-1.31}$ & $1.88^{+0.58}_{-0.61}$ \\ 
0430+052 & - & $0.51^{+0.11}_{-0.1}$ & - \\ 
0458-020 & - & $3.22^{+1.69}_{-1.34}$ & - \\ 
0528+134 & - & $7.82^{+2.73}_{-2.28}$ & $3.5^{+1.4}_{-0.9}$ \\ 
0605-085 & - & $1.68^{+1.13}_{-0.77}$ & - \\ 
0716+714 & $2.44^{+1.05}_{-1.12}$ & $2.01^{+0.52}_{-0.7}$ & - \\ 
0736+017 & $2.53^{+0.35}_{-0.36}$ & $0.95^{+0.22}_{-0.18}$ & - \\ 
0814+425 & - & $0.79^{+0.58}_{-0.22}$ & - \\ 
0851+202 & $5.65^{+1.02}_{-0.8}$ & $3.16^{+0.61}_{-0.71}$ & $1.68^{+0.37}_{-0.44}$ \\ 
0954+658 & - & $4.7^{+1.41}_{-1.45}$ & - \\ 
1055+018 & $6.67^{+0.94}_{-1.21}$ & $5.35^{+1.17}_{-0.87}$ & $1.37^{+1.29}_{-0.74}$ \\ 
1101+384 & $0.97^{+0.25}_{-0.46}$ & - & - \\ 
1156+295 & $7.15^{+2.55}_{-2.46}$ & $3.35^{+0.95}_{-0.85}$ & - \\ 
1222+216 & $2.15^{+1.11}_{-1.05}$ & $1.18^{+0.23}_{-0.26}$ & - \\ 
1226+023 & $3.35^{+0.68}_{-0.73}$ & $5.47^{+2.21}_{-1.93}$ & $2.13^{+0.55}_{-0.53}$ \\ 
1253-055 & $15.35^{+3.95}_{-2.31}$ & $9.87^{+4.37}_{-2.98}$ & $2.21^{+0.66}_{-0.45}$ \\ 
1308+326 & - & $2.3^{+0.59}_{-0.48}$ & - \\ 
1335-127 & $3.31^{+1.59}_{-1.2}$ & $3.81^{+1.09}_{-0.9}$ & - \\ 
1413+135 & - & $0.51^{+0.17}_{-0.12}$ & - \\ 
1418+546 & - & $0.31^{+0.18}_{-0.16}$ & - \\ 
1502+106 & - & $3.32^{+1.03}_{-0.84}$ & - \\ 
1510-089 & $3.68^{+1.37}_{-0.83}$ & $2.99^{+0.84}_{-0.68}$ & - \\ 
1611+343 & - & $1.54^{+0.8}_{-0.53}$ & - \\ 
1633+382 & - & $4.53^{+1.27}_{-0.61}$ & - \\ 
1641+399 & $4.94^{+1.57}_{-2.98}$ & $4.23^{+0.71}_{-0.64}$ & - \\ 
1642+690 & - & $1.7^{+1.01}_{-0.93}$ & - \\ 
1730-130 & $8.47^{+2.07}_{-2.2}$ & $5.24^{+2.56}_{-1.97}$ & $2.13^{+0.8}_{-0.79}$ \\ 
1741-038 & - & $5.83^{+1.58}_{-1.27}$ & $0.95^{+0.99}_{-0.41}$ \\ 
1749+096 & $5.53^{+1.38}_{-1.19}$ & $3.65^{+1.15}_{-1.05}$ & $1.01^{+0.24}_{-0.21}$ \\ 
1803+784 & - & $2.28^{+0.39}_{-0.3}$ & - \\ 
1807+698 & - & $0.25^{+0.11}_{-0.11}$ & - \\ 
2121+053 & - & $1.54^{+0.7}_{-0.4}$ & - \\ 
2200+420 & $1.89^{+0.49}_{-0.31}$ & $1.84^{+0.58}_{-0.38}$ & $0.66^{+0.22}_{-0.18}$ \\ 
2223-052 & - & $3.25^{+0.87}_{-0.69}$ & - \\ 
2230+114 & $10.59^{+2.96}_{-2.99}$ & $7.52^{+1.4}_{-1.58}$ & - \\ 
2251+158 & $16.47^{+2.51}_{-2.05}$ & $9.96^{+3.36}_{-2.9}$ & $9.69^{+2.92}_{-1.79}$ \\ 

\hline 

\end{longtable}

\twocolumn

\end{appendix}

%

\end{document}